# Online calibration of a linear micro tomosynthesis scanner

Piroz Bahar, David Nguyen, Muyang Wang, Dumitru Mazilu, Eric E. Bennett, Han Wen*

Laboratory of Imaging Physics, Biochemistry and Biophysics Center, Division of Intramural Research, National Heart, Lung and Blood Institute, National Institutes of Health, Bethesda, Maryland, USA;

* Corresponding author: Han Wen, National Heart, Lung and Blood Institute, National Institutes of Health, Building 10, Room B1D523, 10 Center Drive, MSC1061, Bethesda, MD 20892, USA. Email: wenh@nhlbi.nih.gov

**Abstract**: In a linear tomosynthesis scanner designed for imaging histologic samples of several centimeter size at 10 µm resolution, the mechanical instability of the scanning stage (±10 µm) exceeded the resolution of the image system, making it necessary to determine the trajectory of the stage for each scan to avoid blurring and artifacts in the images that would arise from the errors in the geometric information used in 3D reconstruction. We present a method for online calibration by attaching a layer of randomly dispersed micro glass beads or calcium particles to the bottom of the sample stage. The marker layer was easy to produce and proven effective in the calibration procedure.

# 1. INTRODUCTION

*Adaptation of linear tomosynthesis to microscopy of pathology samples*

Tomosynthesis is an imaging technique that involves taking a series of x-ray images at different projection angles and reconstructing those images to form a stack of cross-sectional images at a range of depths[1]. It has the advantage of being able to resolve overlapping structures at different depths with relatively simple hardware. In the field of breast imaging, tomosynthesis has helped overcome challenges such as tissue superposition and false positive rates, thereby becoming widely accepted today[2,3]. Examples of other clinical applications besides mammography have also been published[4–6]. A particular type of linear tomosynthesis scan involves moving the sample down a straight line parallel to a flat panel detector[7–10]. It is related in concept to an early form of tomography called planigraphy[11]. In the field of non-destructive testing it is sometimes called laminography, because it is particularly suited to scanning objects of flat shapes[7,8,10]. While this type of tomosynthesis was developed for materials testing and security screening, we were the first to realize that it was also advantageous for pathology samples that have flat shapes, such as standard paraffin embedding cassettes and tissue in petri dishes. Therefore, we adapted it in the form of a linear micro tomosynthesis scanner for microscopy of tissue samples[12,13]. It provided 7 to 10 µm resolution with good soft-tissue contrast in a 15-minute scan, which was an effective scouting tool to guide thin sectioning, staining and light microscopy[12,13]. The scanner is an x-ray cabinet system (Fig. 1). The x-ray hardware is stationary, while the sample is carried on a motorized stage and scanned across a wide cone beam. The scanning geometry allows flat samples to be laid close to the x-ray source, as a way to increase the geometric magnification and the beam intensity through the sample. These factors then help to improve resolution and signal-to-noise ratio. Image reconstruction in tomosynthesis requires knowing the projection matrix from the 3D coordinates in the sample space to the 2D coordinates in the projection image. In practice, the geometry of the x-ray hardware and the sample deviate from the ideal design, and thus a calibration procedure is used to determine the deviations.

*What is new about the present calibration method*

Several recent methods have been published for the calibration of clinical tomosynthesis imaging systems[14–19]. These are based on offline scans of calibration phantoms. What is different about the current scanner is that the mechanical instability of the motorized sample stage (±10 µm) exceeded the resolution of the imaging system, and additionally the instability fluctuated with the scan range and speed. It meant that the geometric deviations were not repeatable, and needed to be calibrated online for each scan.

As a brief review of the literature, there is wealth of calibration methods for cone-beam computed tomography (CT) and tomosynthesis imaging in diverse applications[14–49]. They can broadly be separated into offline phantom-based calibration methods and online phantom-less methods[39]. Graetz[47] and Jiang et al.[17] provided representative literature reviews of offline methods. For offline methods, the geometry of the imaging system is determined with dedicated scans of calibration phantoms. Calibration phantoms contain distinct positional markers, such as radio-opaque beads[15,18,34,35,41,42,45,49] or wires[44]. Methods employing radio-active markers are used to calibrate SPECT imaging systems[20,24]. Some phantom-based methods do not require knowledge of the positions of the markers in the phantom, as long as the assembly of markers remain a rigid unit during calibration scans[23,25,35,36,45,47]. These self-calibration methods work well for micro CT systems that employ microscopic phantoms. The second category of methods are online phantom-less calibration. These are based on the data from the sample scan itself, and do not require separate calibration scans or calibration phantoms [21,29–32,39,40]. Phantom-less methods construct cost functions from the sample data that reflect the amount of errors in the geometric parameters. The cost functions are iteratively minimized leading to the true geometric parameters.

To our knowledge, there were no published methods for online calibration of tomosynthesis scanners. Existing online calibration methods are phantom-less methods for CT scans of the full 180° or 360° projection angles. They either utilize the symmetry properties of a 360° circular scan[21,29,39], or rely on image blurring and artifacts arising from geometric misalignment in full-angle CT scans[30,32]. Compared to CT scans, tomosynthesis collects data from a limited range of projection angles. Image artifacts are inherent due to the lack of full projection angles. Therefore, it can be unreliable to measure geometric errors from image artifacts or blurring, unless the sample contains an abundance of edges or points. Most of our pathology samples were soft tissue specimens that lack such sharp features.

From the above considerations, we created a hybrid method of online phantom-based calibration for tomosynthesis scans. The phantom is a dispersed layer of point makers (glass micro beads) placed at a sufficient distance below the sample platform, such that any tomosynthesis artifacts from the markers are confined in depth and do not interfere with the sample images. Our calibration algorithm has similarities to that of Stevens et al. for a circular-trajectory tomosynthesis scanner[23]. We operate on unfiltered back-projection images of the marker layer at and near focus. From the distribution of local deviations, we directly calculate the geometric parameters in a parametrized version of the sample stage movement. The parameters were then fed into the tomosynthesis reconstruction to improve the quality of the sample reconstruction.

## 2. MATERIALS AND METHODS

*Scanner hardware geometry and scan procedure*

Referring to Fig. 1, the scanner consists of a stationary micro-focus x-ray tube and a stationary flat panel detector. In between the two, a motorized sample stage travels in a plane that is ideally parallel to the image plane in the detector, and in directions that are ideally parallel to the rows or columns of the pixel matrix of the detector. The geometric magnification factor of the system is generally between 10x and 16x depending on the sample thickness[13].

During a sample scan, the stage moves across the x-ray cone-beam at a constant speed in a straight line. The detector acquires projection images at a constant frame rate throughout the scan. As the sample traverses the cone beam, each point in the sample experiences x-rays of continually changing directions, which is equivalent to acquiring projection images from a range of angles. The maximum tomosynthesis angle is thus the angle subtended by the cone beam. A typical setting of scan range and speed for samples in standard tissue-embedding cassettes was 24.5 mm at 0.0272 mm/s, 15 min scan time.

Weighted filtered back-projection was used to reconstruct z-stacks of cross-sectional images at user specified ranges of depth. A parametrized version of the sample stage movement was part of the input for the image reconstruction, which is described below.

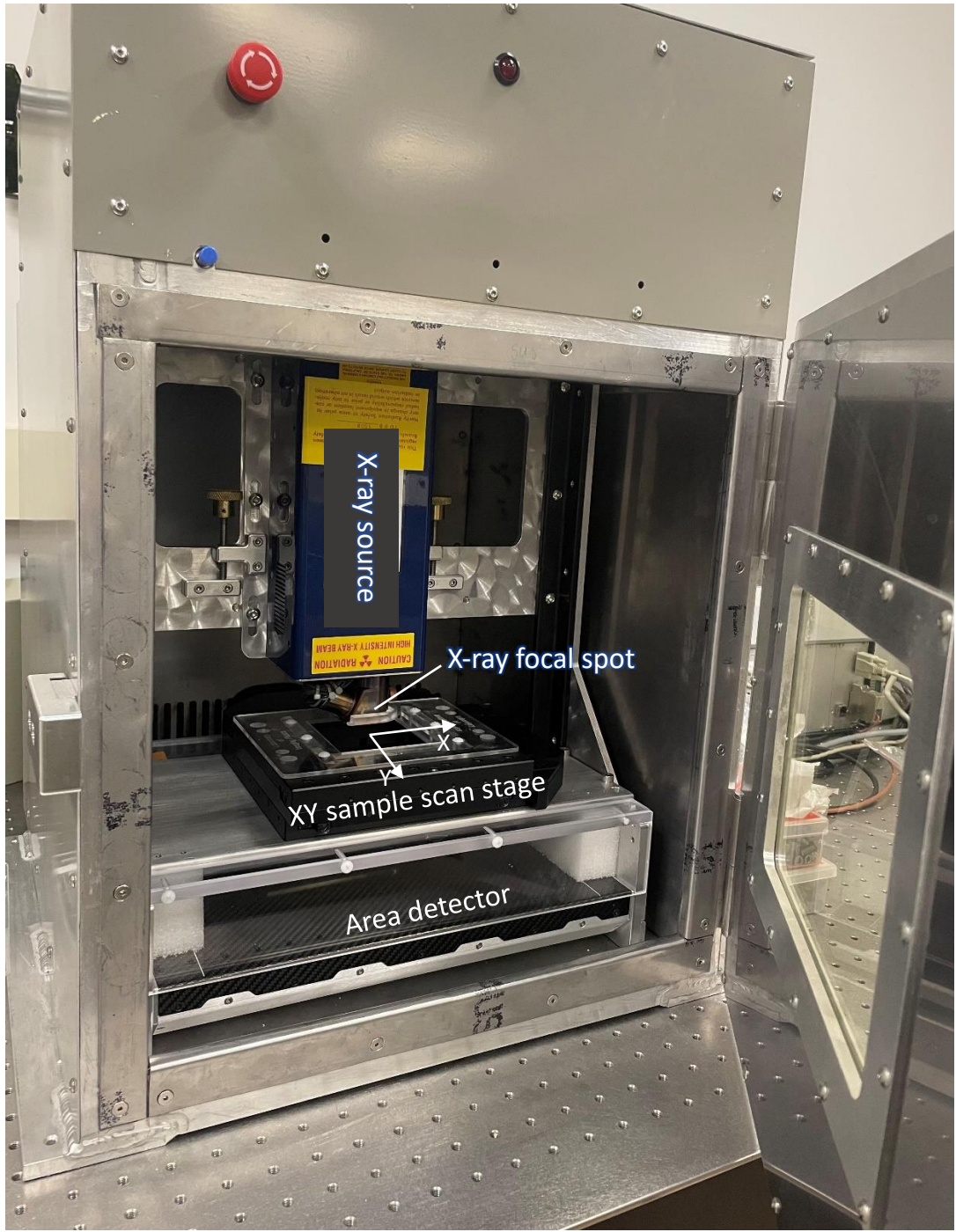

 **Figure 1** The x-ray micro tomosynthesis scanner with the radiation enclosure. The outer dimensions of the enclosure are 45x52x66 cm (width x depth x height). The motorized sample stage moves nominally in the horizontal plane. The sample is scanned horizontally across the vertical cone beam in either x or y direction. Flat samples are usually scanned in a horizontal plane near the x-ray focal spot to maximize the photon flux density through the sample and the geometric magnification for greater contrast-to-noise ration and resolution.

*Theoretical basis of the calibration method*

The basic assumption is that the sample and the sample stage move as a rigid body during the scan. Referring to Fig. 2, geometric calibration involves the transformation between two coordinate systems. The first is the stationary scanner coordinate system (x, y, z), with its origin at the x-ray focal spot, the Z axis pointing perpendicular to the detector, and the X and Y axes parallel to the rows and columns of the detector image matrix. The second is the sample coordinate system ($x_s$, $y_s$, $z_s$), which travels with the sample stage. It is defined to coincide with the stationary scanner coordinate system at the midpoint of the scan. The ideal scan is a translational movement along the X axis. Generally small deviations from the ideal movement have six degrees of freedom. These are functions of the travel distance $l$ of the stage, including the 3D positional errors $d_x(l)$, $d_y(l)$ and $d_z(l)$, and small rotations of the sample stage represented by the three Euler angles $\alpha(l)$, $\beta(l)$ and $\gamma(l)$. By definition of the coordinate systems, the deviations are zero at the midpoint of the scan. We define $l = 0$ at the midpoint. To the leading order in the deviations, the general form of the transformation matrix from the sample to the scanner coordinates is

$$\begin{pmatrix} x \\ y \\ z \\ 1 \end{pmatrix} \approx \begin{pmatrix} 1 & -\gamma(l) & -\beta(l) & l + d_x(l) \\ \gamma(l) & 1 & -\alpha(l) & d_y(l) \\ \beta(l) & \alpha(l) & 1 & d_z(l) \\ 0 & 0 & 0 & 1 \end{pmatrix} \begin{pmatrix} x_s \\ y_s \\ z_s \\ 1 \end{pmatrix}. \qquad (1)$$

The deviations can be expanded in Taylor series of the travel distance $l$. Based on experimental data explained below, only the linear terms in $l$ were significant relative to the resolution of our image system. Thus, the deviations are simplified to the form

$$\begin{pmatrix} d_x(l) \\ d_y(l) \\ d_z(l) \\ \alpha(l) \\ \beta(l) \\ \gamma(l) \end{pmatrix} \approx \begin{pmatrix} \delta_x \\ \delta_y \\ \delta_z \\ \omega_x \\ \omega_y \\ \omega_y \end{pmatrix} l + O(l^2) \qquad (2)$$

It is further simplified by dropping the term $\delta_x l$ in the X direction, which is a scaling of the reconstructed images and does not affect image resolution. The $\delta_y$ and $\delta_z$ represent lateral deviations of the scan direction from the ideal X direction, and the $\omega$'s represent the roll, pitch and yaw of the sample stage.

The projection matrix from the 3D sample coordinates to the 2D coordinates of the detector is given by

$$\begin{pmatrix} x_p \\ y_p \end{pmatrix} = \frac{SID}{z} \begin{pmatrix} x \\ y \end{pmatrix}, \qquad (3)$$

where *SID* is the perpendicular distance from the focal spot to the image plane in the detector. Substituting Eq. (2) into (1), and the results into (3), keeping only first order terms of the deviations and $l$, the 2D coordinates in the image plane are given by the expression

$$\begin{pmatrix} x_p \\ y_p \end{pmatrix} \approx \frac{SID}{z_s} \begin{pmatrix} x_s + l + O(\delta, \omega)l \\ y_s + \delta_y l - z_s \omega_x l + x_s \omega_z l - y_s \frac{\delta_z}{z_s} l - x_s y_s \frac{\omega_y}{z_s} l - y_s^2 \frac{\omega_x}{z_s} l \end{pmatrix}. \qquad (4)$$

Equation (4) shows that for any fixed point in the sample, its projected 2D trajectory on the image plane may deviate from the ideal horizontal direction. The in-plane deviation angle varies with the 3D coordinates of the point in the sample coordinate system, which is given by the expression:

$$\delta_p(x_s, y_s, z_s) \approx \delta_y - z_s \omega_x + x_s \omega_z - y_s \frac{\delta_z}{z_s} - x_s y_s \frac{\omega_y}{z_s} - y_s^2 \frac{\omega_x}{z_s}. \qquad (5)$$

Equation (5) is the theoretical basis for the present calibration method. The next section explains how Eq.(5) was observed experimentally for each sample scan, and how it was used to determine the parameters of the sample stage movement.

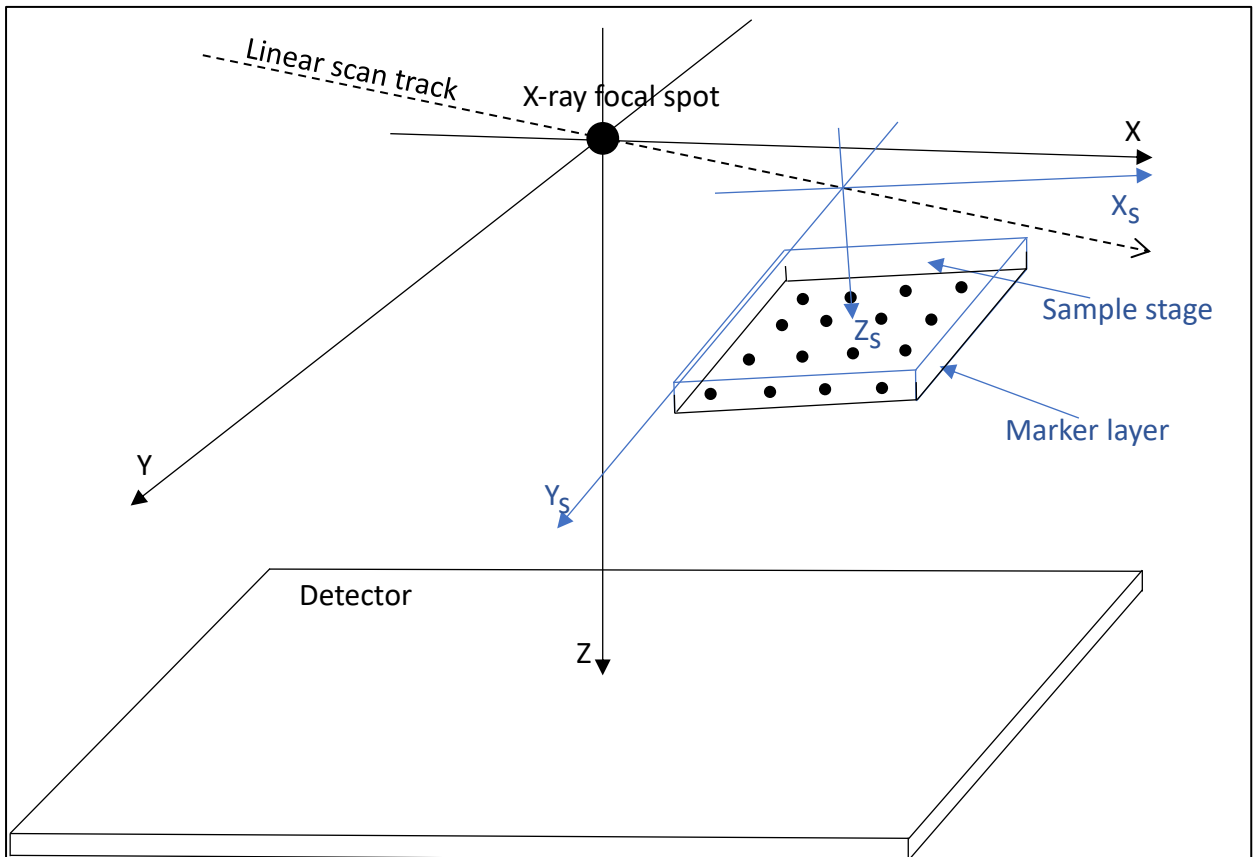

**Figure 2** The definition of the sample coordinate system ($X_s$, $Y_s$, $Z_s$) and the stationary scanner coordinate system (X, Y, Z). The sample coordinate system is attached rigidly to the sample stage and moves with it during the scan. The two coordinate systems coincide with each other at the mid time point of the scan. The glass microbead layer serving as fiducial markers for online geometric calibration is rigidly attached to the sample stage, typically at 1 cm below the sample. The scan track may not be perfectly aligned with the X-axis and the sample stage may rotate slightly during the scan. An exaggerated version of such misalignment is illustrated.

*Measurement of the geometric parameters from a layer of dispersed markers*

We used three protocols to create a dispersed layer of markers to be attached to the sample stage. One was dispersing glass microbeads of 20 µm diameter on the surface of a 1 mm thick acrylic plate by suspending both the microbeads and plate in water. The microbeads covered a square of 10 cm size on the plate. The plate was rigidly attached to the sample stage at a level of 10 mm below the bottom surface. A second method was gluing a sheet of printer paper to the surface of an acrylic plate, whereby the cellulose fibers of the paper were the markers. The third method was dispersing a layer of hydroxyapatite powder on the acrylic plate and covering it with a 10 cm wide adhesive tape. Both the glass microbead and hydroxyapatite powder procedures produced a satisfactory marker layer, although the hydroxyapatite procedure was simpler. The printer paper had overly strong x-ray contrasts that made calculations unreliable.

After a sample scan, a z-stack of images centered at the depth of the marker layer was reconstructed with weighted filtered back-projection, assuming ideal scan movement. The z-stack covered a z range of 1 to 2 mm at 5 µm increment with an in-plane pixel size of 7 µm. In slightly off-focus images of the marker layer, the markers defocused into short line segments (Fig. 3). The line segments were straight but tilted from the horizontal direction. The tilt represents an in-plane deviation of the projected trajectory of the bead from the X axis. The straightness of the line segments meant that the deviations of the scan movement were approximately linear with respect to the scan position. This was the basis for the linear approximation expressed by Eq.(2). The in-plane deviation angle varied with the location of the beads, which is illustrated in Fig. 3. This was anticipated by Eq.(5).

The next step was to obtain the in-plane deviation angle $\delta_p$ as a function of the location of the markers. The data could then be used in Eq.(5) to estimate the geometric parameters of the sample stage movement. For this purpose, the entire marker layer was divided into a rectangular grid of 60 to 200 grid points. At each grid point, a small z-stack of 400 images covering a volume of 1.4x1.4 mm in-plane and 2 mm depth was reconstructed around the z depth of the marker layer. The reconstruction was repeated while assuming a range of in-plane tilt angles of the scan trajectory, resulting in multiple small z-stacks. The in-plane deviation angle $\delta_p$ at each grid point was determined as the scan trajectory angle that maximized the sharpness of the markers in the local z-stack. For each z-stack, marker sharpness was measured by the standard deviation of the image pixel values in the slice where the markers come into focus. That slice was determined as the one with the highest standard deviation value. At each grid point, among the multiple z-stacks of different scan trajectory angles, the scan trajectory angle giving the maximal marker sharpness was the sought-after deviation angle $\delta_p$ for that grid point. This process was repeated for all grid points, and collectively provided the function $\delta_p(x_s, y_s, z_s)$ within the marker layer. Examples of the function is shown in the surface plots of Fig. 4.

The measured function $\delta_p(x_s, y_s, z_s)$ from the marker layer was fitted to Eq.(5) with polynomial fitting. The marker layer had a single depth $z_s$, which reduced Eq.(5) to a 2$^{nd}$ order function of the in-plane coordinates $x_s$ and $y_s$. The fitted coefficients of the 6 terms in Eq.(5) then provided all the geometric parameters of the sample stage movement, including $\delta_y, \delta_z, \omega_x, \omega_y, \omega_z$.

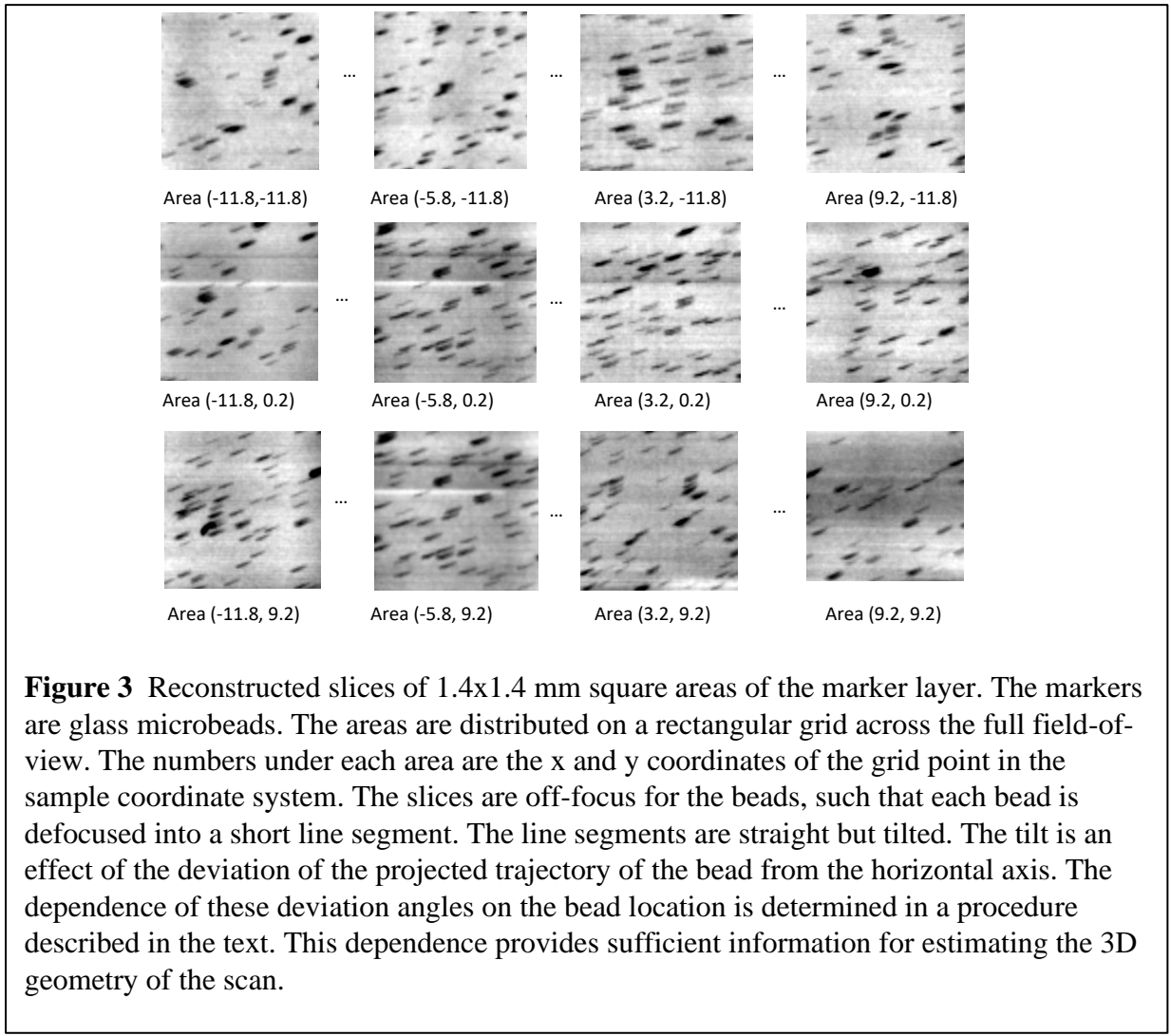

**Figure 3** Reconstructed slices of 1.4x1.4 mm square areas of the marker layer. The markers are glass microbeads. The areas are distributed on a rectangular grid across the full field-of-view. The numbers under each area are the x and y coordinates of the grid point in the sample coordinate system. The slices are off-focus for the beads, such that each bead is defocused into a short line segment. The line segments are straight but tilted. The tilt is an effect of the deviation of the projected trajectory of the bead from the horizontal axis. The dependence of these deviation angles on the bead location is determined in a procedure described in the text. This dependence provides sufficient information for estimating the 3D geometry of the scan.

## 3. RESULTS

As described in the Methods section, the key intermediate step of the calibration process is to determine the in-plane deviation angles of the marker trajectories from the prescribed scan direction, as a function of the location of the markers. Examples of the results for three different scan settings are shown Figs. 4A-4C. These figures are surface plots of the measured in-plane deviation angle as a function of the $x_s$ and $y_s$ coordinates of the markers. Variation of these plots illustrates that the geometric parameters varied with the scan ranges and speeds.

Table I summarizes the estimated geometric parameters for the three different scan settings. Also tabulated are the errors of the calculated marker positions over the course of the scan, before and after applying the geometric parameters, and the root-mean-square (RMS) of the deviation angles over the field of view before and after the calibration. In all three scan settings, the residual errors were less than 7.5% of the un-calibrated levels. The residual position errors after the calibration were all below the system resolution of 7 µm. The residual deviation angles of the marker trajectories as a function of the marker location are plotted in Fig. 4D-4E.

The effect of the calibration procedure on the reconstructed images is demonstrated in two samples. The first is a plexiglass plate coated with a dispersed layer of hydroxyapatite particles, and the second a paraffin-embedded mouse heart sample containing calcified atherosclerotic plaques in the aorta. The results are summarized in Figs. 5 and 6, respectively. In both cases, we observed an increase in the resolution of the reconstructed images after the calibration was applied.

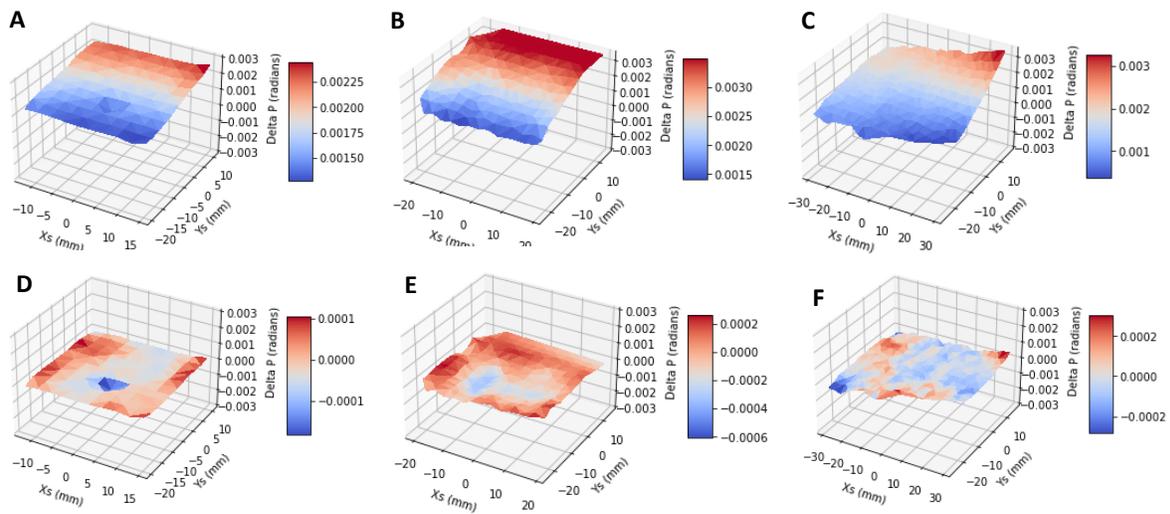

**Figure 4** Surface plots of the measured in-plane deviation angles of the marker trajectories relative to the ideal horizontal direction, as a function of the marker position in the sample coordinate system. Plot **A** is from a scan of the typical speed of 0.0272 mm/s and range of 24 mm, **B** a scan of half the typical speed and range, and **C** a scan of twice the typical speed and range. Plots **D-E** are the residual deviation angles after subtracting the fit of the model of the scan geometry described by Eq.(5) in Methods. The RMS of the residual deviation angles is 7.5% or less of the RMS of the original deviation angles.

Table I. List of the calibrated geometric parameters for different scan settings, and errors of the marker positions before and after calibration. The errors include the RMS of the residual deviation angles of the scan trajectories of the markers, and the residual error of the marker positions relative to the estimated positions.

|  | Typical scan speed and range | Half scan speed and range | Double scan speed and range | Typical scan setting #2 trial |
|---|---|---|---|---|
| $\delta_y$ (radians) | 1.76E-03 | 3.98E-03 | 1.58E-03 | 1.92E-03 |
| $\delta_z$ (radians) | -1.04E-03 | -1.25E-03 | -1.71E-03 | -1.66E-03 |
| $\omega_x$ (radians/mm) | -6.88E-06 | 3.18E-05 | -2.28E-06 | -5.17E-06 |
| $\omega_y$ (radians/mm) | -1.18E-05 | -1.23E-05 | -4.14E-05 | -3.57E-05 |
| $\omega_z$ (radians/mm) | -6.94E-07 | 1.68E-06 | -2.30E-07 | -6.69E-07 |
| RMS of the deviation angles without calibration (radians) | 1.83E-03 | 2.80E-03 | 1.66E-03 | 1.98E-03 |
| RMS of the residual deviation angles with calibration (radians) | 6.66E-05 | 1.71E-04 | 1.18E-04 | 1.35E-04 |
| Error range of marker positions without calibration (µm) | 45.75 | 35.01 | 83.18 | 50.04 |
| Error range of marker positions with calibration (µm) | 1.66 | 2.14 | 5.92 | 3.40 |

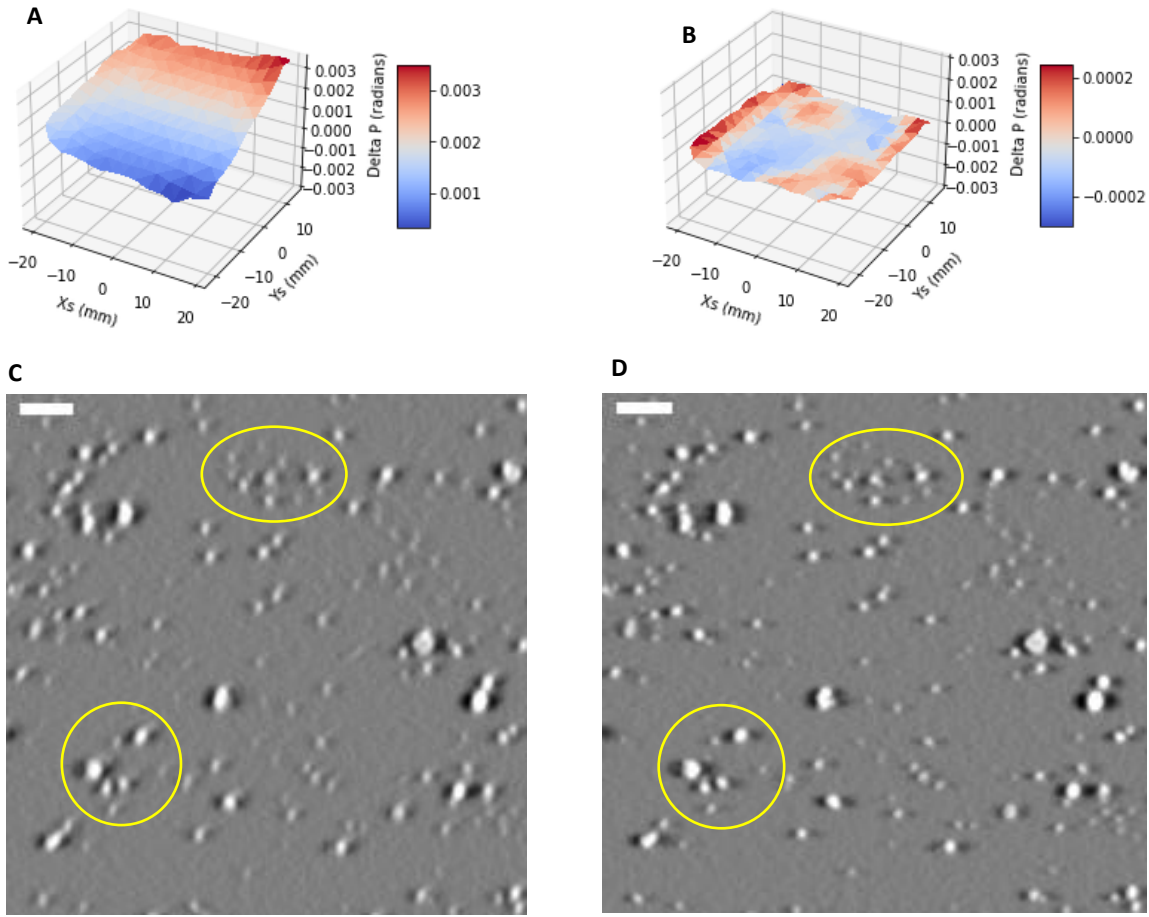

**Figure 5** The effect of geometric calibration on the reconstructed image of a layer of hydroxyapatite particles. **A** is the surface plot of the deviation angles of the scan trajectories of the markers as a function of the marker position. The layer of the glass bead markers was fixed at 10 mm below the sample stage. **B** is the residual of the measurement in **A** after subtracting the fitted model of geometric calibration. **C** and **D** are the reconstructed images of the hydroxyapatite particles before and after applying the calibration. The yellow circles highlight the visible changes of the resolution of the particles. The scale bars are 160 µm.

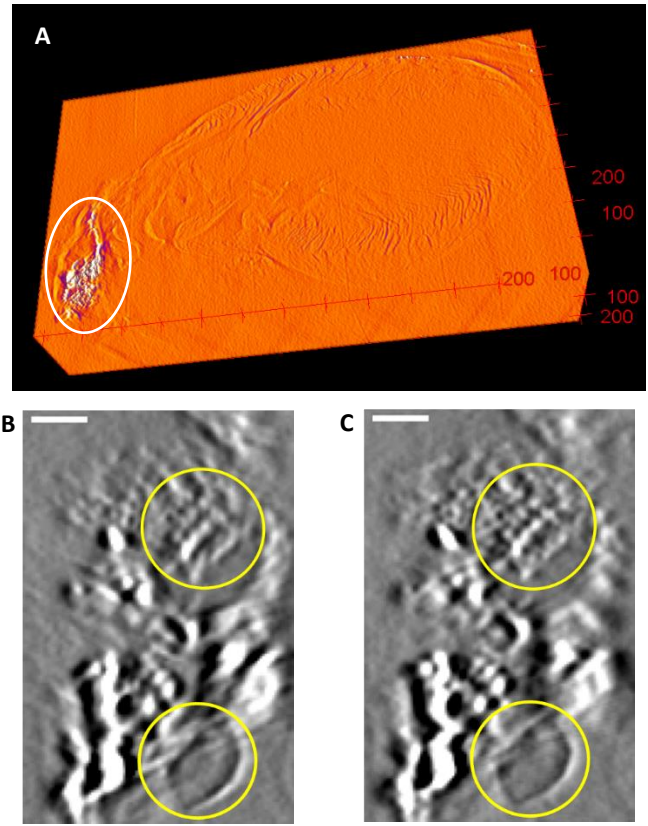

**Figure 6**. **A** is the reconstructed 3D stack of a paraffin-embedded mouse heart sample including an aorta segment. The white oval highlights a calcified lesion in the aorta. **B** and **C** are magnified views of the calcified lesion before and after applying geometric calibration. The white scale bar is equivalent to 160 µm. The yellow circles highlight the internal grain texture of the calcification and an air bubble.

## 4. DISCUSSION

Since the tomosynthesis scanner presented in this paper was designed to image tissue samples at 10 µm resolution, it was necessary to determine the precise movement and rotation of the sample stage every time a scan was performed, such that image reconstruction was corrected for minor mechanical instabilities. These instabilities were not repeatable from scan to scan (Table I) and were significant enough to caused artifacts and blurriness in the resulting images if not corrected. Although the calibration method presented here required a layer of point markers to be attached to the bottom of the sample stage, it did not require any specific pattern of distribution for the markers, as long as they were dispersed throughout the imaging field of view. In practice, a layer of dispersed hydroxyapatite powder on a thin acrylic plate was simple to make, and effective for soft tissue samples. A necessary condition is that the marker layer is rigidly attached to the sample stage, such that the two move as a rigid body. Another key point of the method is that the vertical separation of the sample and the marker layer is sufficient to avoid mutual interference. For tissue samples in standard embedding cassettes, a 10 mm separation was found to be sufficient.

## References

1. Dobbins, J.T.; Godfrey, D.J. Digital X-Ray Tomosynthesis: Current State of the Art and Clinical Potential. *Phys. Med. Biol.* **2003**, *48*, R65–R106, doi:10.1088/0031-9155/48/19/R01.
2. Vedantham, S.; Karellas, A.; Vijayaraghavan, G.R.; Kopans, D.B. Digital Breast Tomosynthesis: State of the Art. *Radiology* **2015**, *277*, 663–684, doi:10.1148/radiol.2015141303.
3. Chong, A.; Weinstein, S.P.; McDonald, E.S.; Conant, E.F. Digital Breast Tomosynthesis: Concepts and Clinical Practice. *Radiology* **2019**, *292*, 1–14, doi:10.1148/radiol.2019180760.
4. Gomi, T.; Nakajima, M.; Fujiwara, H.; Takeda, T.; Saito, K.; Umeda, T.; Sakaguchi, K. Comparison between Chest Digital Tomosynthesis and CT as a Screening Method to Detect Artificial Pulmonary Nodules: A Phantom Study. *Br J Radiol* **2012**, *85*, e622–e629, doi:10.1259/bjr/12643098.
5. Blum, A.; Noël, A.; Regent, D.; Villani, N.; Gillet, R.; Gondim Teixeira, P. Tomosynthesis in Musculoskeletal Pathology. *Diagnostic and Interventional Imaging* **2018**, *99*, 423–441, doi:10.1016/j.diii.2018.05.001.
6. Machida, H.; Yuhara, T.; Tamura, M.; Ishikawa, T.; Tate, E.; Ueno, E.; Nye, K.; Sabol, J.M. Whole-Body Clinical Applications of Digital Tomosynthesis. *RadioGraphics* **2016**, *36*, 735–750, doi:10.1148/rg.2016150184.
7. Zhou, J.; Maisl, M.; Reiter, H.; Arnold, W. Computed Laminography for Materials Testing. *Appl Phys Lett* **1996**, *68*, 3500–3502.
8. Gondrom, S.; Zhou, J.; Maisl, M.; Reiter, H.; Kröning, M.; Arnold, W. X-Ray Computed Laminography: An Approach of Computed Tomography for Applications with Limited Access. *Nuclear Engineering and Design* **1999**, *190*, 141–147, doi:10.1016/S0029-5493(98)00319-7.
9. Gao, H.; Zhang, L.; Chen, Z.; Xing, Y.; Xue, H.; Cheng, J. Straight-Line-Trajectory-Based X-Ray Tomographic Imaging for Security Inspections: System Design, Image Reconstruction and Preliminary Results. *IEEE Transactions on Nuclear Science* **2013**, *60*, 3955–3968, doi:10.1109/TNS.2013.2274481.
10. O'Brien, N.S.; Boardman, R.P.; Sinclair, I.; Blumensath, T. Recent Advances in X-Ray Cone-Beam Computed Laminography. *J. X-Ray Sci. Technol.* **2016**, *24*, 691–707, doi:10.3233/XST-160581.

11. Howes, W.E. Planigraphy—Its Application to Thoracic Diagnosis. *Radiology* **1939**, *32*, 556–566, doi:10.1148/32.5.556.
12. Wen, H.; Martinez, A.M.; Miao, H.X.; Larsen, T.C.; Nguyen, C.P.; Bennett, E.E.; Moorse, K.P.; Yu, Z.X.; Remaley, A.T.; Boehm, M.; et al. Correlative Detection of Isolated Single and Multi-Cellular Calcifications in the Internal Elastic Lamina of Human Coronary Artery Samples. *Scientific Reports* **2018**, *8*, 10978.
13. Nguyen, D.T.; Larsen, T.C.; Wang, M.; Knutsen, R.H.; Yang, Z.; Bennett, E.E.; Mazilu, D.; Yu, Z.-X.; Tao, X.; Donahue, D.R.; et al. X-Ray Microtomosynthesis of Unstained Pathology Tissue Samples. *Journal of Microscopy* **2021**, *283*, 9–20, doi:10.1111/jmi.13003.
14. Wang, X.; Mainprize, J.G.; Kempston, M.P.; Mawdsley, G.E.; Yaffe, M.J. Digital Breast Tomosynthesis Geometry Calibration. In Proceedings of the Medical Imaging 2007: Physics of Medical Imaging; SPIE, March 17 2007; Vol. 6510, pp. 1118–1128.
15. Li, X.; Zhang, D.; Liu, B. A Generic Geometric Calibration Method for Tomographic Imaging Systems with Flat-Panel Detectors—A Detailed Implementation Guide. *Medical Physics* **2010**, *37*, 3844–3854, doi:10.1118/1.3431996.
16. Miao, H.; Wu, X.; Zhao, H.; Liu, H. A Phantom-Based Calibration Method for Digital x-Ray Tomosynthesis. *Journal of X-Ray Science and Technology* **2012**, *20*, 17–29, doi:10.3233/XST-2012-0316.
17. Jiang, C.; Zhang, N.; Gao, J.; Hu, Z. Geometric Calibration of a Stationary Digital Breast Tomosynthesis System Based on Distributed Carbon Nanotube X-Ray Source Arrays. *PLOS ONE* **2017**, *12*, e0188367, doi:10.1371/journal.pone.0188367.
18. Choi, C.J.; Vent, T.L.; Acciavatti, R.J.; Maidment, A.D.A. Geometric Calibration for a Next-Generation Digital Breast Tomosynthesis System Using Virtual Line Segments. In Proceedings of the Medical Imaging 2018: Physics of Medical Imaging; SPIE, March 9 2018; Vol. 10573, pp. 89–98.
19. Chang, C.-H.; Ni, Y.-C.; Huang, S.-Y.; Hsieh, H.-H.; Tseng, S.-P.; Tseng, F.-P. A Geometric Calibration Method for the Digital Chest Tomosynthesis with Dual-Axis Scanning Geometry. *PLOS ONE* **2019**, *14*, e0216054, doi:10.1371/journal.pone.0216054.
20. Gullberg, G.T.; Tsui, B.M.W.; Crawford, C.R.; Ballard, J.G.; Hagius, J.T. Estimation of Geometrical Parameters and Collimator Evaluation for Cone Beam Tomography. *Medical Physics* **1990**, *17*, 264–272, doi:10.1118/1.596505.
21. Azevedo, S.G.; Schneberk, D.J.; Fitch, J.P.; Martz, H.E. Calculation of the Rotational Centers in Computed Tomography Sinograms. *IEEE Transactions on Nuclear Science* **1990**, *37*, 1525–1540, doi:10.1109/23.55866.
22. Noo, F.; Clackdoyle, R.; Mennessier, C.; White, T.A.; Roney, T.J. Analytic Method Based on Identification of Ellipse Parameters for Scanner Calibration in Cone-Beam Tomography. *Phys.Med.Biol.* **2000**, *45*, 3489–3508.
23. Stevens, G.M.; Saunders, R.; Pelc, N.J. Alignment of a Volumetric Tomography System. *Medical Physics* **2001**, *28*, 1472–1481, doi:10.1118/1.1382609.
24. Beque, D.; Nuyts, J.; Bormans, G.; Suetens, P.; Dupont, P. Characterization of Pinhole SPECT Acquisition Geometry. *IEEE Transactions on Medical Imaging* **2003**, *22*, 599–612, doi:10.1109/TMI.2003.812258.
25. von Smekal, L.; Kachelriess, M.; Stepina, E.; Kalender, W.A. Geometric Misalignment and Calibration in Cone-Beam Tomography. *Med Phys* **2004**, *31*, 3242–3266, doi:10.1118/1.1803792.

26. Cho, Y.; Moseley, D.J.; Siewerdsen, J.H.; Jaffray, D.A. Accurate Technique for Complete Geometric Calibration of Cone-Beam Computed Tomography Systems. *Medical Physics* **2005**, *32*, 968–983, doi:10.1118/1.1869652.
27. Yang, K.; Kwan, A.L.C.; Miller, D.F.; Boone, J.M. A Geometric Calibration Method for Cone Beam CT Systems. *Medical Physics* **2006**, *33*, 1695–1706, doi:10.1118/1.2198187.
28. Hoppe, S.; Noo, F.; Dennerlein, F.; Lauritsch, G.; Hornegger, J. Geometric Calibration of the Circle-plus-Arc Trajectory. *Phys. Med. Biol.* **2007**, *52*, 6943–6960, doi:10.1088/0031-9155/52/23/012.
29. Panetta, D.; Belcari, N.; Guerra, A.D.; Moehrs, S. An Optimization-Based Method for Geometrical Calibration in Cone-Beam CT without Dedicated Phantoms. *Phys. Med. Biol.* **2008**, *53*, 3841–3861, doi:10.1088/0031-9155/53/14/009.
30. Kyriakou, Y.; Lapp, R.M.; Hillebrand, L.; Ertel, D.; Kalender, W.A. Simultaneous Misalignment Correction for Approximate Circular Cone-Beam Computed Tomography. *Phys Med Biol* **2008**, *53*, 6267–6289, doi:10.1088/0031-9155/53/22/001.
31. Patel, V.; Chityala, R.N.; Hoffmann, K.R.; Ionita, C.N.; Bednarek, D.R.; Rudin, S. Self-Calibration of a Cone-Beam Micro-CT System. *Med Phys* **2009**, *36*, 48–58, doi:10.1118/1.3026615.
32. Kingston, A.; Sakellariou, A.; Varslot, T.; Myers, G.; Sheppard, A. Reliable Automatic Alignment of Tomographic Projection Data by Passive Auto-Focus. *Medical Physics* **2011**, *38*, 4934–4945, doi:10.1118/1.3609096.
33. Li, X.; Zhang, D.; Liu, B. Sensitivity Analysis of a Geometric Calibration Method Using Projection Matrices for Digital Tomosynthesis Systems. *Medical Physics* **2011**, *38*, 202–209, doi:10.1118/1.3524221.
34. Wu, D.; Li, L.; Zhang, L.; Xing, Y.; Chen, Z.; Xiao, Y. Geometric Calibration of Cone-Beam CT with a Flat-Panel Detector. In Proceedings of the 2011 IEEE Nuclear Science Symposium Conference Record; October 2011; pp. 2952–2955.
35. Sawall, S.; Knaup, M.; Kachelrieß, M. A Robust Geometry Estimation Method for Spiral, Sequential and Circular Cone-Beam Micro-CT. *Medical Physics* **2012**, *39*, 5384–5392, doi:10.1118/1.4739506.
36. Gross, D.; Heil, U.; Schulze, R.; Schoemer, E.; Schwanecke, U. Auto Calibration of a Cone-Beam-CT. *Medical Physics* **2012**, *39*, 5959–5970, doi:10.1118/1.4739247.
37. Wicklein, J.; Kunze, H.; Kalender, W.A.; Kyriakou, Y. Image Features for Misalignment Correction in Medical Flat-Detector CT. *Medical Physics* **2012**, *39*, 4918–4931, doi:10.1118/1.4736532.
38. Ladikos, A.; Wein, W. Geometric Calibration Using Bundle Adjustment for Cone-Beam Computed Tomography Devices. In Proceedings of the Medical Imaging 2012: Physics of Medical Imaging; SPIE, March 3 2012; Vol. 8313, pp. 814–819.
39. Meng, Y.; Gong, H.; Yang, X. Online Geometric Calibration of Cone-Beam Computed Tomography for Arbitrary Imaging Objects. *IEEE Transactions on Medical Imaging* **2013**, *32*, 278–288, doi:10.1109/TMI.2012.2224360.
40. Ben Tekaya, I.; Kaftandjian, V.; Buyens, F.; Sevestre, S.; Legoupil, S. Registration-Based Geometric Calibration of Industrial X-Ray Tomography System. *IEEE Transactions on Nuclear Science* **2013**, *60*, 3937–3944, doi:10.1109/TNS.2013.2279675.
41. Xu, M.; Zhang, C.; Liu, X.; Li, D. Direct Determination of Cone-Beam Geometric Parameters Using the Helical Phantom. *Phys. Med. Biol.* **2014**, *59*, 5667–5690, doi:10.1088/0031-9155/59/19/5667.

42. Zechner, A.; Stock, M.; Kellner, D.; Ziegler, I.; Keuschnigg, P.; Huber, P.; Mayer, U.; Sedlmayer, F.; Deutschmann, H.; Steininger, P. Development and First Use of a Novel Cylindrical Ball Bearing Phantom for 9-DOF Geometric Calibrations of Flat Panel Imaging Devices Used in Image-Guided Ion Beam Therapy. *Phys. Med. Biol.* **2016**, *61*, N592–N605, doi:10.1088/0031-9155/61/22/N592.
43. Zhou, K.; Huang, Y.; Meng, X.; Li, Z.; Li, S.; Yang, K.; Ren, Q. A New Method for Cone-Beam Computed Tomography Geometric Parameters Estimation. *Journal of Computer Assisted Tomography* **2016**, *40*, 639–648, doi:10.1097/RCT.0000000000000393.
44. Jacobson, M.W.; Ketcha, M.D.; Capostagno, S.; Martin, A.; Uneri, A.; Goerres, J.; Silva, T.D.; Reaungamornrat, S.; Han, R.; Manbachi, A.; et al. A Line Fiducial Method for Geometric Calibration of Cone-Beam CT Systems with Diverse Scan Trajectories. *Phys. Med. Biol.* **2018**, *63*, 025030, doi:10.1088/1361-6560/aa9910.
45. Li, G.; Luo, S.; You, C.; Getzin, M.; Zheng, L.; Wang, G.; Gu, N. A Novel Calibration Method Incorporating Nonlinear Optimization and Ball-Bearing Markers for Cone-Beam CT with a Parameterized Trajectory. *Medical Physics* **2019**, *46*, 152–164, doi:10.1002/mp.13278.
46. Nguyen, V.; Sanctorum, J.G.; Van Wassenbergh, S.; Dirckx, J.J.J.; Sijbers, J.; De Beenhouwer, J. Geometry Calibration of a Modular Stereo Cone-Beam X-Ray CT System. *Journal of Imaging* **2021**, *7*, 54, doi:10.3390/jimaging7030054.
47. Graetz, J. Auto-Calibration of Cone Beam Geometries from Arbitrary Rotating Markers Using a Vector Geometry Formulation of Projection Matrices. *Phys. Med. Biol.* **2021**, *66*, 075013, doi:10.1088/1361-6560/abe75f.
48. Moon, S.; Choi, S.; Jang, H.; Shin, M.; Roh, Y.; Baek, J. Geometry Calibration and Image Reconstruction for Carbon-Nanotube-Based Multisource and Multidetector CT. *Phys. Med. Biol.* **2021**, *66*, 165005, doi:10.1088/1361-6560/ac16c1.
49. Duan, X.; Cai, J.; Ling, Q.; Huang, Y.; Qi, H.; Chen, Y.; Zhou, L.; Xu, Y. Knowledge-Based Self-Calibration Method of Calibration Phantom by and for Accurate Robot-Based CT Imaging Systems. *Knowledge-Based Systems* **2021**, *229*, 107343, doi:10.1016/j.knosys.2021.107343.